\documentclass[sigconf]{acmart}
\AtBeginDocument{%
  }

\copyrightyear{2026}
\acmYear{2026}
\setcopyright{cc}
\setcctype{by}
\acmConference[CHI EA '26]{Extended Abstracts of the 2026 CHI Conference on Human Factors in Computing Systems}{April 13--17, 2026}{Barcelona, Spain}
\acmBooktitle{Extended Abstracts of the 2026 CHI Conference on Human Factors in Computing Systems (CHI EA '26), April 13--17, 2026, Barcelona, Spain}
\acmDOI{10.1145/3772363.3798931}
\acmISBN{979-8-4007-2281-3/2026/04}

\acmISBN{978-1-4503-XXXX-X/2018/06}

\usepackage{hyperref}
\usepackage{cleveref}
\usepackage{subcaption}

\begin{document}

\title[Exploring Interactive LLM Scaffolding to Support Learning Engagement]{From Passive Consumption to Active Interaction: Exploring Interactive LLM Scaffolding to Support Learning Engagement}


\author{Zixin Chen}
\orcid{0000-0001-8507-4399}
\affiliation{%
  \institution{The Hong Kong University of Science and Technology}
  \city{Hong Kong}
  \country{China}
}
\email{zchendf@connect.ust.hk}

\author{Haotian Li}
\orcid{0000-0001-9547-3449}
\affiliation{%
  \institution{Microsoft Research Asia}
  \city{Beijing}
  \country{China}
}
\email{haotian.li@microsoft.com}

\author{Zhe Liu}
\orcid{0000-0002-1904-9045}
\affiliation{%
  \institution{University of British Columbia}
  \city{Vancouver, British Columbia}
  \country{Canada}
}
\email{zheliu92@cs.ubc.ca}

\author{Huamin Qu}
\orcid{0000-0002-3344-9694}
\affiliation{%
  \institution{The Hong Kong University of Science and Technology}
  \city{Hong Kong}
  \country{China}
}
\email{huamin@cse.ust.hk}

\author{Xing Xie}
\orcid{0009-0009-3257-3077}
\affiliation{%
  \institution{Microsoft Research Asia}
  \city{Beijing}
  \country{China}
}
\email{xingx@microsoft.com}

\renewcommand{\shortauthors}{Trovato et al.}
\newcommand{\haotian}[1]{\textcolor{teal}{[Haotian: #1]}}

\begin{abstract}


Large Language Models (LLMs) are increasingly used as learning companions, providing scaffolded explanations, hints, or step-by-step guidance. However, in current LLM-based learning scenarios, scaffolded content is primarily consumed passively, offering limited support for active learner engagement. Learning science research suggests that effective educational scaffolding depends not only on what support is provided, but also on how learners engage with it. In this work, we explore whether embedding lightweight interactive components into LLM-generated scaffolding responses can promote learning-oriented engagement and improve short-term learning outcomes. We evaluated this approach through a within-subjects laboratory study (N=8). Results provide initial evidence that interactive scaffolding increases learners' perceived engagement and attentional focus, while supporting short-term learning performance. We conclude with design implications for integrating interaction into LLM-generated scaffolding to support active learning engagement.

\end{abstract}

\ccsdesc[500]{Human-centered computing~Interactive systems and tools}


\keywords{Interactive scaffolding, large language model, learning engagement}


\maketitle

\section{Introduction}



Large Language Models (LLMs) are increasingly integrated into everyday activities, emerging as a common form of personal support~\cite{li2024personal}. In educational contexts, a growing number of learners have begun to treat LLMs as personal tutors for real-time learning support, using them to explain questions, assist with exercises, and even recommend personalized learning materials~\cite{lieb2024student, neumann2024llm, chen2025cograder, chen2025unmasking, sharma2025role}.


In response to this trend, recent research has explored the development of LLM-powered intelligent tutoring systems that better approximate effective human teaching practices by incorporating principles from the learning sciences~\cite{park2024empowering, stamper2024enhancing}. Among various tutoring strategies, scaffolding has emerged as one of the most widely adopted approaches, as its emphasis on adaptive and auxiliary support naturally aligns with learner–LLM interactions~\cite{tian2024theory}. Prior work has demonstrated that LLMs are capable of breaking down complex ideas, providing partial guidance, and structuring solution processes in ways that respond to learners' needs~\cite{liu2024scaffolding, shao2025unlocking, goslen2025llm, chen2024stugptviz, ma2025dbox, chen2026question}.




While these practices closely reflect long-established educational theories of scaffolding, their delivery is often constrained by standard chatbot interfaces, which primarily rely on static textual responses~\cite{adeshola2024opportunities}. As a result, LLM-based scaffolding is typically consumed through reading model outputs~\cite{rahman2023chatgpt, baidoo2023education}, in contrast to traditional classroom settings where learners are encouraged to actively engage with instructional materials—for example, by filling in missing steps, annotating key statements, or completing partially worked solutions~\cite{deslauriers2019measuring}. Consequently, current LLM-based scaffolding offers limited support for learners to \emph{interact with} scaffolded content itself, beyond passively receiving it~\cite{bhattacharjee2024understanding, malik2025scaffolding}.

\begin{figure*}[!t]
  \includegraphics[width=\textwidth]{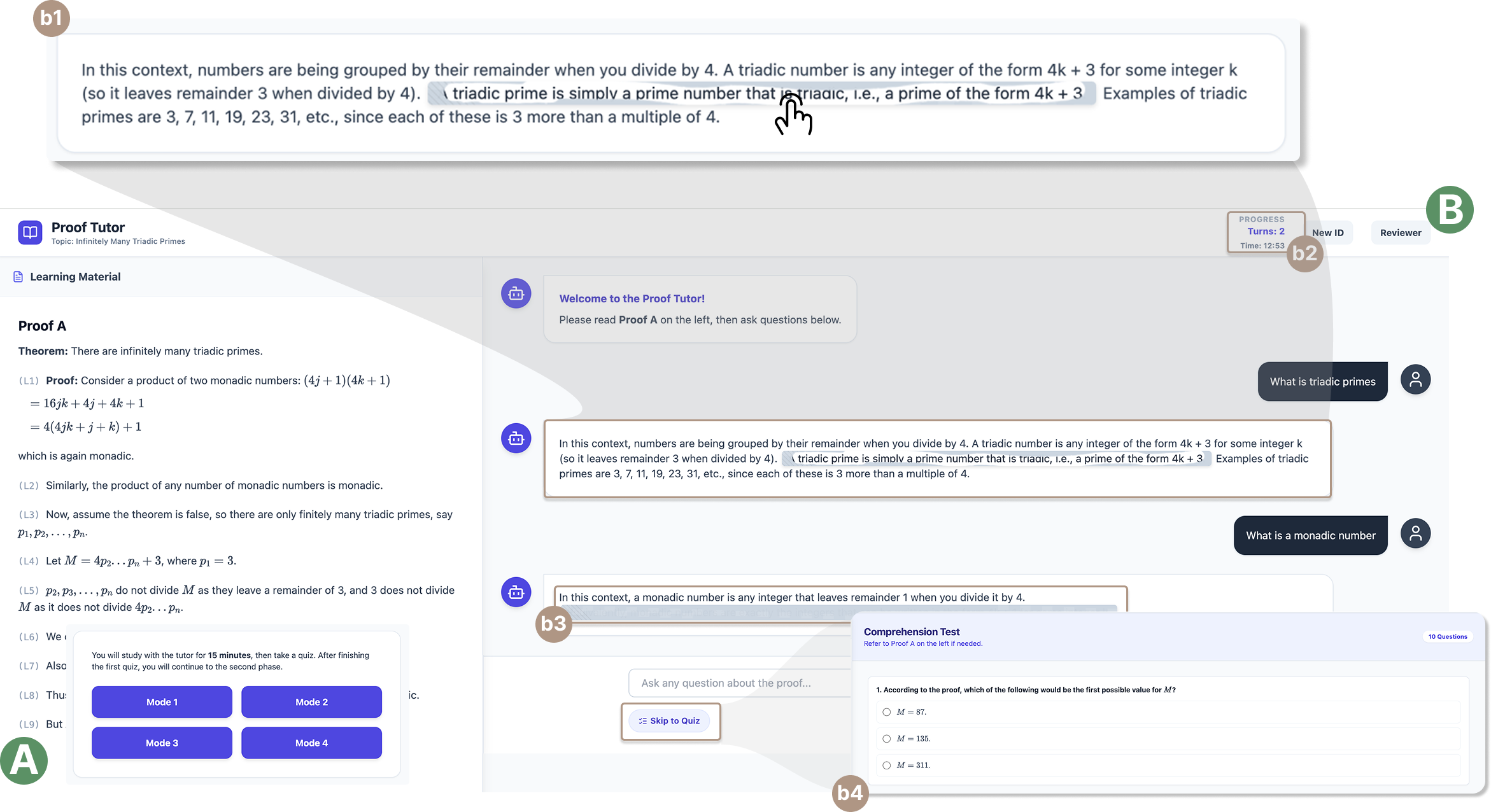}
  \caption{Overview of the LLM-based tutoring prototype.
(A) The Learning Material Zone presents the target proof and allows learners to select the scaffolding condition.
(B) The Chatbot Tutor Zone supports question-driven interaction with the LLM; scaffolded content is revealed via scratch-off interaction in the interactive condition and shown as plain text in the non-interactive condition.}
  \label{fig:teaser}
\end{figure*}

This limitation is critical because decades of learning sciences research suggest that effective learning outcomes depend not only on the presence of scaffolding, but also on how learners engage with instructional materials~\cite{nortvig2018literature, xu2023student, lin2017study}. According to the widely adopted ICAP framework~\cite{chi2014icap}, learner engagement ranges from passive to active, constructive, and interactive, depending on learners' observable actions. In current LLM-based learning scenarios, engagement remains skewed toward passive consumption. While constructive and interactive engagement may occasionally occur when learners generate new ideas or ask follow-up questions through dialogue~\cite{scarlatos2025training,shahriar2024confused,chi2014icap}, the primary educational content in LLM-based learning—the scaffolded content within LLM responses—is most often consumed passively, for example through reading generated explanations~\cite{lieb2024student, techawitthayachinda2024automatic}. Active engagement, which involves interacting or working directly with the provided instructional materials, remains weakly supported in conventional conversational interfaces~\cite{lieb2024student, hao2025student}. As a result, the key challenge for future AI learning scaffolding lies not in \textit{what} LLMs can generate, but in \textit{how} scaffolded content is presented to guide learners' attention and promote active leaarning engagement.

In this work, we empirically investigate whether embedding lightweight interactive mechanisms into LLM-generated scaffolding responses can promote learners' active engagement and benefit short-term learning in conversation-based learning settings. 
We treat interaction as a means to deliberately guide learners' attention to scaffolded content, inspired by recent interface designs that support targeted and dynamic interactions with LLM-generated content~\cite{shen2025interaction}.
Specifically, we design an interactive LLM-based learning system as a design probe. In this system, scaffolded content within LLM responses (e.g., concept explanations) is initially masked and can only be revealed through a simple scratch-off interaction, requiring learners to actively engage with the content before accessing it.

We evaluate this design through a within-subjects laboratory study (N = 8) using a college-level number theory proof comprehension task adapted from prior educational research~\cite{hodds2014self}. Participants learned two mathematical proofs with LLM support under two conditions: interactive scaffolding and non-interactive (static) scaffolding. Learning outcomes were assessed using corresponding comprehension quizzes, while learners' engagement experiences and perceptions were examined through post-task surveys. In addition, we conducted think-aloud interview studies to elicit participants' interaction design ideas for enriching interactive scaffolding. Together, this work provides initial empirical evidence and design insights showing how lightweight interaction, serving as a design probe, can guide learners' attention to scaffolded content and support learning-oriented engagement in LLM-based learning.

\section{Methods}

\begin{figure*}[!t]
  \centering
  \includegraphics[width=\linewidth]{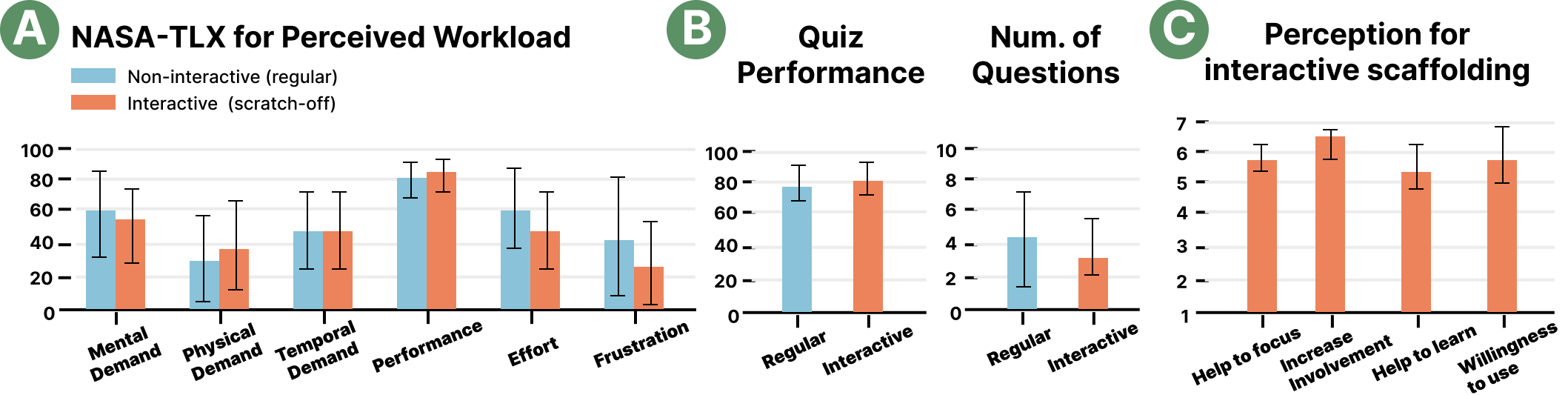}
  \caption{(A) NASA-TLX workload ratings across conditions. (B) Proof comprehension quiz performance across conditions and number of questions asked during the learning phase. (C) Perceived helpfulness and engagement of interactive scaffolding (7-point Likert). }
  \label{fig:statistics}
\end{figure*}

This study aims to empirically examine whether adding lightweight interaction to LLM-generated scaffolding can influence learners' engagement experiences and short-term learning outcomes. To this end, we designed an interactive LLM tutoring prototype as a design probe and conducted a controlled within-subject laboratory study comparing interactive and non-interactive scaffolding conditions.


\subsection{Prototype Design}



We developed an LLM-based tutoring prototype that supports both \emph{interactive} and \emph{non-interactive} scaffolding variants in the LLM's responses. It allows learners to study mathematical proofs with LLM support either through a standard chatbot interface or through an interactive mode in which scaffolded content is revealed via a scratch-off interaction.


The interface consists of two main zones: a \textit{Learning Material Zone} (\cref{fig:teaser}-A) and a \textit{Chatbot Tutor Zone} (\cref{fig:teaser}-B). In the Learning Material Zone, learners view the target proof and select the scaffolding condition. In the Chatbot Tutor Zone, learners ask questions about the proof and receive LLM responses. Under the interactive condition, scaffolded content within the responses (e.g., definitions or explanatory guidance) is initially masked and must be actively revealed through a mouse-based scratch-off interaction (\cref{fig:teaser}-b1, b3). Under the non-interactive condition, the same content is presented as plain text. When the learning phase ends (\cref{fig:teaser}-b2), the Chatbot Tutor Zone switches to a quiz interface (\cref{fig:teaser}-b4), and learners can no longer interact with the LLM.

The system uses the GPT-5.2 model with tailored prompts that encourage established scaffolding practices, such as hints, explanations, and step-by-step guidance, following prior scaffolding theory~\cite{van2010scaffolding}. To support interactive scaffolding, the LLM explicitly marks scaffolded content segments in its responses, which are then detected by a client-side script to enable scratch-off interaction. 
To assess the reliability of LLM-identified scaffolding segments, we conducted a pilot study with two participants prior to the main experiment. Two authors independently annotated 36 scaffolded segments identified by the LLM using definitions from prior scaffolding literature~\cite{van2010scaffolding}. Among these segments, 75\% were recognized as scaffolding by both annotators and 83.33\% by at least one annotator, indicating sufficient reliability for use in the main study. Prompt templates are provided in the Appendix.

\subsection{Experimental Design}

\noindent\textbf{Learning Task.}
The learning task was adapted from prior educational research on proof comprehension~\cite{hodds2014self}. We selected two college-level number theory proofs (Proof A and Proof C from~\cite{hodds2014self}), each accompanied by a corresponding set of multiple-choice comprehension questions developed in the original study. The proofs were presented line by line within the system interface. The quizzes assessed learners' understanding of proof structure, logical dependencies, and key reasoning steps, and were used as a measure of short-term learning outcomes. Detailed materials are provided in the Appendix.

\noindent\textbf{Design and Conditions.}
We employed a within-subjects design to compare two scaffolding conditions: \textit{interactive scaffolding} and \textit{non-interactive scaffolding}. Each participant experienced both conditions, learning one proof under each condition. To mitigate ordering effects, the assignment of proof (A vs.\ C) to scaffolding condition and the order in which participants encountered the two proofs were counterbalanced across participants. This resulted in four counterbalancing sequences, with two participants assigned to each sequence. Across both conditions, the instructional content and LLM behavior were held constant, with the only difference being whether scaffolded content in the LLM’s responses was presented with interactive masking or as plain text.

\noindent\textbf{Procedure.}
The IRB-approved study procedure lasted approximately 60 minutes and consisted of five phases:

\begin{figure*}[!t]
  \centering
  \includegraphics[width=0.775\linewidth]{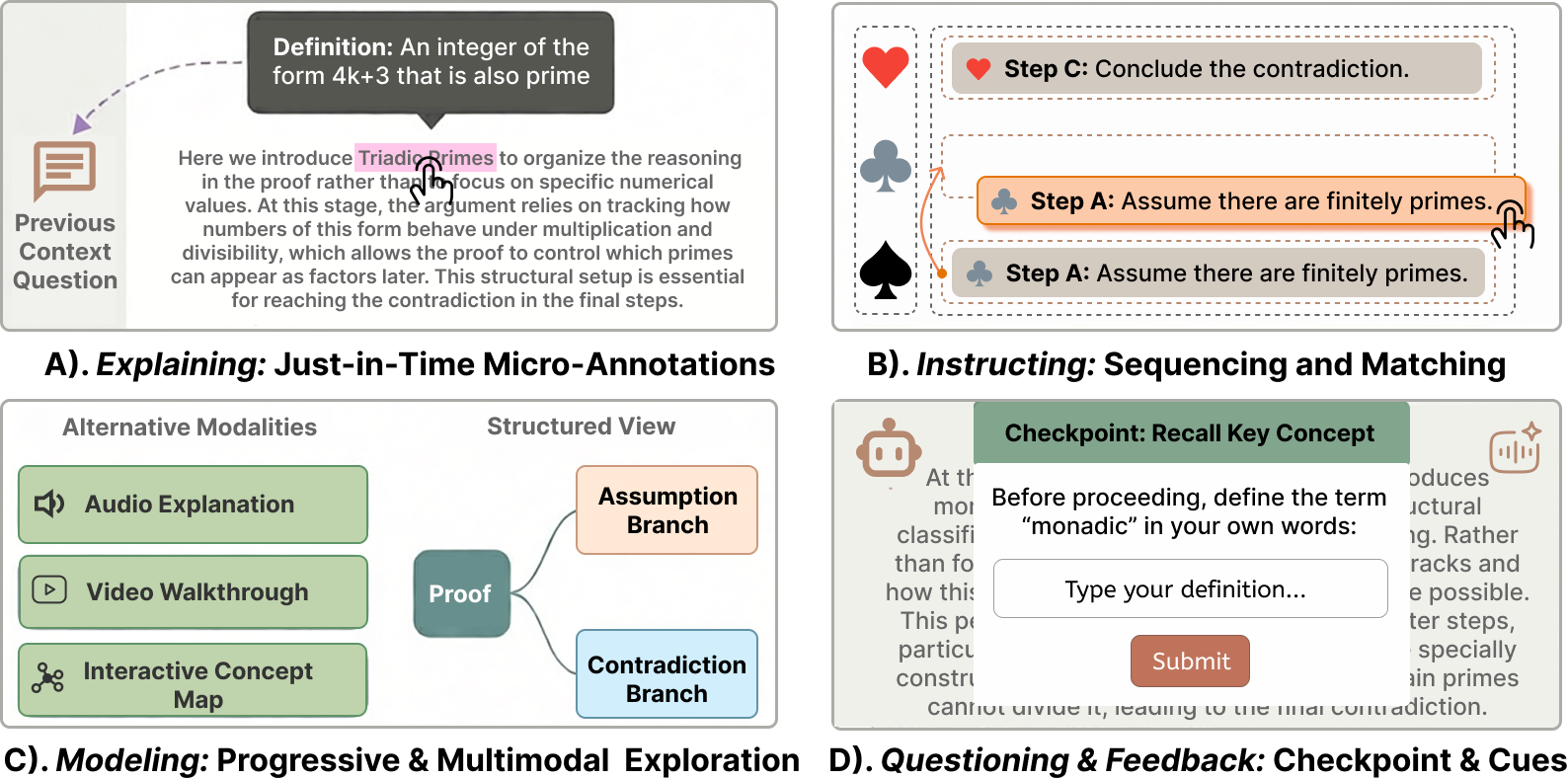}
  \vspace{-6pt}
  \caption{Interaction design ideas proposed by participants during the think-aloud brainstorming. Designs are analyzed and organized by \emph{scaffolding means}, illustrating how different interaction forms align with distinct pedagogical roles of scaffolding.}
  \label{fig:interaction}
\end{figure*}

\textbf{1)}. After reviewing the study information and providing informed consent, participants first completed a brief exposure phase. They were given three minutes to skim both Proof~A and Proof~C to verify that they were not already familiar with either proof. Participants were explicitly instructed that this phase was not intended for learning, but only to confirm unfamiliarity with the materials.

\textbf{2)}. Participants then completed two ``learning--quiz'' cycles, one for each proof. First, participants studied a proof with LLM support for up to 15 minutes under the assigned scaffolding condition (interactive or non-interactive). They worked through the proof line by line and could freely ask questions to the LLM. Participants could also end the learning phase early and proceed directly to the quiz.

\textbf{3)}. Participants then completed a comprehension quiz corresponding to the studied proof. Each quiz consisted of 10 multiple-choice questions and was completed under a 10-minute time limit.


\textbf{4)}. Next, Participants proceeded to the second ``learning--quiz'' cycle, in which they studied the remaining proof under the alternate scaffolding condition, followed by its corresponding comprehension quiz with the same time constraints.

\textbf{5)}. Finally, participants completed a post-study questionnaire assessing their experiences with both scaffolding conditions, focusing on perceived engagement, attention allocation, and perceived usefulness for learning. Participants then took part in a 10-minute think-aloud interview to brainstorm interaction design ideas for LLM-generated scaffolding.


\noindent\textbf{Participants.}
We recruited eight participants via email invitations at the authors' institution. All participants were computer science PhD students with undergraduate-level mathematics backgrounds. They reported prior experience learning mathematical proofs but no prior familiarity with the two target theorems used in this study. The study was conducted in a controlled laboratory setting.

\section{Findings}
We report quantitative results from post-study questionnaires and proof comprehension quizzes, complemented by participants' open-ended feedback (N=8).

\noindent\textbf{Workload and Usability. }
Overall, participants perceived the interactive scaffolding interface as usable and no more demanding than the non-interactive chatbot condition. NASA-TLX ratings~\cite{hart1988development} indicate that introducing interaction did not increase perceived workload: the interactive condition showed slightly lower overall workload (M=49.8, SD=14.9) than the non-interactive condition (M=52.8, SD=17.3) (\cref{fig:statistics}-A). Across subscales, \textit{Effort} and \textit{Frustration} were lower with interactive scaffolding, while \textit{Physical Demand} was slightly higher due to the additional mouse-based interaction. Usability was rated positively, with high SUS scores (M=81.6, SD=8.1; range 70--95), indicating that the interactive interface was easy to learn and use. It is possible that interaction structured learners' attention and pacing when engaging with scaffolded content, reducing perceived effort and frustration despite the added physical interaction.

\noindent\textbf{Learning Outcomes and Learning Behaviors. }
Interactive scaffolding showed a positive trend in learning outcomes (\cref{fig:statistics}-B). Participants achieved slightly higher quiz scores with interactive scaffolding (M=80.0\%, SD=11.95) than with non-interactive scaffolding (M=77.5\%, SD=11.65). Learning behaviors also differed directionally: participants asked fewer questions under the interactive condition (M=3.75, SD=1.75) than under the non-interactive condition (M=4.38, SD=3.02), suggesting that interaction may support understanding with fewer conversational turns.

\noindent\textbf{Perceived Engagement and Helpfulness. }
Survey responses indicate that participants perceived interactive scaffolding as supportive of learning-oriented engagement (7-point Likert; \cref{fig:statistics}-C). Participants agreed that the interaction helped them focus on relevant proof content (M=5.75, SD=0.46), increased active involvement compared to plain text (M=6.13, SD=0.64), and was helpful for learning overall (M=5.50, SD=0.76). Participants also reported a clear preference for the interactive format over a standard chatbot (M=5.75, SD=1.04).

\section{Design Insights from Think-Aloud Interview Study}


To move beyond a single interaction, we conducted a think-aloud interview with a brief co-design component following the lab study. Our goal was to elicit participants' intuitions about how interaction could better support different forms of scaffolding in LLM-based learning. Guided by prior scaffolding literature~\cite{van2010scaffolding}, we organize these bottom-up ideas around six \emph{scaffolding means}—including hinting, instructing, explaining, modeling, feedback, and questioning—and illustrate how each may benefit from distinct interaction designs. In particular, we identify the scratch-off design probe as most naturally aligned with \emph{hinting}, as it delays access to partial guidance while remaining lightweight.

\noindent\textbf{Explaining through Just-in-Time Micro-Annotations.} For scaffolding with an \emph{explanatory} role, our analysis highlights the value of localized, on-demand interactions that clarify meaning without disrupting the learning flow. Across participants (e.g., P1, P4), a recurring pattern was to highlight key terms or phrases in the LLM's response and allow learners to access brief explanations when needed. Common examples included hover-based micro-annotations on highlighted terms, which temporarily reveal concise definitions, contextual reminders, or clarifications before fading back into the main response (\cref{fig:interaction}-A). Participants also envisioned explicitly linking explanations to learners' original questions, for instance through visual anchors pointing back to triggering keywords. Together, these designs frame explanation as just-in-time clarification that supports focused attention within scaffolded content.

\noindent\textbf{Instructing through Sequencing and Matching Interactions.} When scaffolding serves an \emph{instructional} purpose, particularly for step-by-step procedures or worked examples, our analysis highlights interaction designs that require learners to actively organize or reconstruct instructional structure embedded in the LLM's response. Participants (e.g., P2, P5) described designs in which instructional steps are presented out of order and learners must reorder them, match steps to rationales, or drag key elements into appropriate positions (\cref{fig:interaction}-B). These sequencing- and matching-based interactions shift instruction from passive consumption to procedural engagement, prompting learners to reason about order, dependency, and structure as they interact with scaffolded guidance.

\noindent\textbf{Modeling through Progressive and Multimodal Exploration.} For \emph{modeling}-oriented scaffolding, where the goal is to demonstrate expert reasoning or solution processes, our synthesis highlights interaction designs that support progressive exposure and structured exploration of modeled content. Participants proposed expandable worked examples, stepwise reveals, and interactive visual structures such as concept maps that learners can unfold incrementally (\cref{fig:interaction}-C). Participants also suggested complementing with additional modalities, including audio narration or short walkthrough videos, to foreground reasoning flow and emphasis. Together, these interaction forms support modeling by helping learners selectively attend to expert strategies and reasoning patterns without being overwhelmed by complete solutions.

\noindent\textbf{Questioning and Feeding Back through Checkpoints and Signals.} In addition to the interaction directions discussed above, our analysis also connects participants' ideas to the remaining scaffolding means. For example, lightweight pop-up checkpoints naturally align with \emph{questioning} by prompting learners to articulate an answer, while immediate audio or visual cues can support \emph{feedback} by providing timely signals following learners' actions (\cref{fig:interaction}-D). 

Taken together, these insights suggest that interaction design in LLM-based learning should not be treated as a one-size-fits-all enhancement. Instead, different \emph{scaffolding means} call for distinct interaction strategies that guide how learners attend to, engage with, and act upon scaffolded content, pointing toward future work on principled mappings between interaction design and pedagogical intent.

\section{Conclusion}

This work explored integrating lightweight interaction into LLM-generated scaffolding and provide initial empirical evidence and design insights for supporting learner engagement in conversation-based learning. Using a design probe, our results highlight interaction as a promising lever for rethinking how LLM-generated scaffolding can guide learners' attention and support learning engagement and outcomes across contexts. As an exploratory study, our findings should be interpreted as preliminary, grounded in a small-scale, short-term laboratory investigation with a relatively homogeneous participant pool. While the interactive condition was associated with higher perceived engagement and modest short-term comprehension gains, the underlying cognitive mechanisms remain to be further clarified—for instance, whether the effect reflects genuinely constructive engagement or simply slower, more attentive reading induced by the interaction. Future work should incorporate richer behavioral evidence and more diverse learner populations to better understand the robustness and generalizability of these effects.

\bibliographystyle{ACM-Reference-Format}
\bibliography{sample-base}

@String{Computing = "Computing" }

@String{Computer = "{IEEE} Computer" }

@String{Academic = "Academic Press" }

@String{Springer = "Springer-Verlag" }

@inproceedings{park2024empowering,
  title={Empowering personalized learning through a conversation-based tutoring system with student modeling},
  author={Park, Minju and Kim, Sojung and Lee, Seunghyun and Kwon, Soonwoo and Kim, Kyuseok},
  booktitle={Extended Abstracts of the CHI Conference on Human Factors in Computing Systems},
  pages={1--10},
  year={2024}
}

@inproceedings{liu2024scaffolding,
  title={Scaffolding language learning via multi-modal tutoring systems with pedagogical instructions},
  author={Liu, Zhengyuan and Yin, Stella Xin and Lee, Carolyn and Chen, Nancy F},
  booktitle={2024 IEEE conference on artificial intelligence (CAI)},
  pages={1258--1265},
  year={2024},
  organization={IEEE}
}

@article{li2024personal,
  title={Personal llm agents: Insights and survey about the capability, efficiency and security},
  author={Li, Yuanchun and Wen, Hao and Wang, Weijun and Li, Xiangyu and Yuan, Yizhen and Liu, Guohong and Liu, Jiacheng and Xu, Wenxing and Wang, Xiang and Sun, Yi and others},
  journal={arXiv preprint arXiv:2401.05459},
  year={2024}
}

@inproceedings{lieb2024student,
  title={Student interaction with newtbot: An llm-as-tutor chatbot for secondary physics education},
  author={Lieb, Anna and Goel, Toshali},
  booktitle={Extended Abstracts of the CHI Conference on Human Factors in Computing Systems},
  pages={1--8},
  year={2024}
}

@inproceedings{stamper2024enhancing,
  title={Enhancing llm-based feedback: Insights from intelligent tutoring systems and the learning sciences},
  author={Stamper, John and Xiao, Ruiwei and Hou, Xinying},
  booktitle={International Conference on Artificial Intelligence in Education},
  pages={32--43},
  year={2024},
  organization={Springer}
}

@article{neumann2024llm,
  title={An llm-driven chatbot in higher education for databases and information systems},
  author={Neumann, Alexander Tobias and Yin, Yue and Sowe, Sulayman and Decker, Stefan and Jarke, Matthias},
  journal={IEEE Transactions on Education},
  year={2024},
  publisher={IEEE}
}

@article{sharma2025role,
  title={The role of large language models in personalized learning: a systematic review of educational impact},
  author={Sharma, Sahil and Mittal, Puneet and Kumar, Mukesh and Bhardwaj, Vivek},
  journal={Discover Sustainability},
  volume={6},
  number={1},
  pages={1--24},
  year={2025},
  publisher={Springer}
}

@inproceedings{shao2025unlocking,
  title={Unlocking Scientific Concepts: How Effective Are LLM-Generated Analogies for Student Understanding and Classroom Practice?},
  author={Shao, Zekai and Yuan, Siyu and Gao, Lin and He, Yixuan and Yang, Deqing and Chen, Siming},
  booktitle={Proceedings of the 2025 CHI Conference on Human Factors in Computing Systems},
  pages={1--19},
  year={2025}
}

@article{goslen2025llm,
  title={Llm-based student plan generation for adaptive scaffolding in game-based learning environments},
  author={Goslen, Alex and Kim, Yeo Jin and Rowe, Jonathan and Lester, James},
  journal={International journal of artificial intelligence in education},
  volume={35},
  number={2},
  pages={533--558},
  year={2025},
  publisher={Springer}
}

@inproceedings{ma2025dbox,
  title={Dbox: Scaffolding algorithmic programming learning through learner-llm co-decomposition},
  author={Ma, Shuai and Wang, Junling and Zhang, Yuanhao and Ma, Xiaojuan and Wang, April Yi},
  booktitle={Proceedings of the 2025 CHI Conference on Human Factors in Computing Systems},
  pages={1--20},
  year={2025}
}

@article{adeshola2024opportunities,
  title={The opportunities and challenges of ChatGPT in education},
  author={Adeshola, Ibrahim and Adepoju, Adeola Praise},
  journal={Interactive Learning Environments},
  volume={32},
  number={10},
  pages={6159--6172},
  year={2024},
  publisher={Taylor \& Francis}
}

@article{rahman2023chatgpt,
  title={ChatGPT for education and research: Opportunities, threats, and strategies},
  author={Rahman, Md Mostafizer and Watanobe, Yutaka},
  journal={Applied sciences},
  volume={13},
  number={9},
  pages={5783},
  year={2023},
  publisher={MDPI}
}

@article{baidoo2023education,
  title={Education in the era of generative artificial intelligence (AI): Understanding the potential benefits of ChatGPT in promoting teaching and learning},
  author={Baidoo-Anu, David and Ansah, Leticia Owusu},
  journal={Journal of AI},
  volume={7},
  number={1},
  pages={52--62},
  year={2023},
  publisher={{\.I}zmir Academy Association}
}

@inproceedings{bhattacharjee2024understanding,
  title={Understanding the role of large language models in personalizing and scaffolding strategies to combat academic procrastination},
  author={Bhattacharjee, Ananya and Zeng, Yuchen and Xu, Sarah Yi and Kulzhabayeva, Dana and Ma, Minyi and Kornfield, Rachel and Ahmed, Syed Ishtiaque and Mariakakis, Alex and Czerwinski, Mary P and Kuzminykh, Anastasia and others},
  booktitle={Proceedings of the 2024 CHI Conference on Human Factors in Computing Systems},
  pages={1--18},
  year={2024}
}

@inproceedings{tian2024theory,
  title={A theory guided scaffolding instruction framework for LLM-enabled metaphor reasoning},
  author={Tian, Yuan and Xu, Nan and Mao, Wenji},
  booktitle={Proceedings of the 2024 Conference of the North American Chapter of the Association for Computational Linguistics: Human Language Technologies (Volume 1: Long Papers)},
  pages={7731--7748},
  year={2024}
}

@article{malik2025scaffolding,
  title={Scaffolding middle school mathematics curricula with large language models},
  author={Malik, Rizwaan and Abdi, Dorna and Wang, Rose and Demszky, Dorottya},
  journal={British Journal of Educational Technology},
  volume={56},
  number={3},
  pages={999--1027},
  year={2025},
  publisher={Wiley Online Library}
}

@article{nortvig2018literature,
  title={A literature review of the factors influencing e-learning and blended learning in relation to learning outcome, student satisfaction and engagement},
  author={Nortvig, Anne-Mette and Petersen, Anne Kristine and Balle, S{\o}ren Hattesen},
  journal={Electronic Journal of E-learning},
  volume={16},
  number={1},
  pages={pp46--55},
  year={2018}
}

@article{xu2023student,
  title={Student engagement and learning outcomes: an empirical study applying a four-dimensional framework},
  author={Xu, Xiaoming and Shi, Zehua and Bos, Nicolaas A and Wu, Hongbin},
  journal={Medical Education Online},
  volume={28},
  number={1},
  pages={2268347},
  year={2023},
  publisher={Taylor \& Francis}
}

@article{lin2017study,
  title={A study of the effects of digital learning on learning motivation and learning outcome},
  author={Lin, Ming-Hung and Chen, Huang-Cheng and Liu, Kuang-Sheng},
  journal={Eurasia journal of mathematics, science and technology education},
  volume={13},
  number={7},
  pages={3553--3564},
  year={2017},
  publisher={Modestum}
}

@article{chi2014icap,
  title={The ICAP framework: Linking cognitive engagement to active learning outcomes},
  author={Chi, Michelene TH and Wylie, Ruth},
  journal={Educational psychologist},
  volume={49},
  number={4},
  pages={219--243},
  year={2014},
  publisher={Taylor \& Francis}
}

@inproceedings{techawitthayachinda2024automatic,
  title={Automatic Assessment of Active Learning in Online Discussions with Large Language Models},
  author={Techawitthayachinda, Ratrapee and Iriya, Rafael},
  booktitle={International Conference on Artificial Intelligence in Education Technology},
  pages={34--42},
  year={2024},
  organization={Springer}
}

@inproceedings{scarlatos2025training,
  title={Training llm-based tutors to improve student learning outcomes in dialogues},
  author={Scarlatos, Alexander and Liu, Naiming and Lee, Jaewook and Baraniuk, Richard and Lan, Andrew},
  booktitle={International Conference on Artificial Intelligence in Education},
  pages={251--266},
  year={2025},
  organization={Springer}
}

@inproceedings{shahriar2024confused,
  title={“I Am Confused! How to Differentiate Between…?” Adaptive Follow-Up Questions Facilitate Tutor Learning with Effective Time-On-Task},
  author={Shahriar, Tasmia and Matsuda, Noboru},
  booktitle={International Conference on Artificial Intelligence in Education},
  pages={17--30},
  year={2024},
  organization={Springer}
}

@article{hao2025student,
  title={Student engagement in collaborative learning with AI agents in an LLM-empowered learning environment: A cluster analysis},
  author={Hao, Zhanxin and Jiang, Jianxiao and Yu, Jifan and Liu, Zhiyuan and Zhang, Yu},
  journal={arXiv preprint arXiv:2503.01694},
  year={2025}
}

@article{hodds2014self,
  title={Self-explanation training improves proof comprehension},
  author={Hodds, Mark and Alcock, Lara and Inglis, Matthew},
  journal={Journal for Research in Mathematics Education},
  volume={45},
  number={1},
  pages={62--101},
  year={2014},
  publisher={National Council of Teachers of Mathematics}
}

@article{van2010scaffolding,
  title={Scaffolding in teacher--student interaction: A decade of research},
  author={Van de Pol, Janneke and Volman, Monique and Beishuizen, Jos},
  journal={Educational psychology review},
  volume={22},
  number={3},
  pages={271--296},
  year={2010},
  publisher={Springer}
}

@incollection{hart1988development,
  title={Development of NASA-TLX (Task Load Index): Results of empirical and theoretical research},
  author={Hart, Sandra G and Staveland, Lowell E},
  booktitle={Advances in psychology},
  volume={52},
  pages={139--183},
  year={1988},
  publisher={Elsevier}
}

@article{deslauriers2019measuring,
  title={Measuring actual learning versus feeling of learning in response to being actively engaged in the classroom},
  author={Deslauriers, Louis and McCarty, Logan S and Miller, Kelly and Callaghan, Kristina and Kestin, Greg},
  journal={Proceedings of the National Academy of Sciences},
  volume={116},
  number={39},
  pages={19251--19257},
  year={2019},
  publisher={National Academy of Sciences}
}

@article{shen2025interaction,
  title={Interaction-Augmented Instruction: Modeling the Synergy of Prompts and Interactions in Human-GenAI Collaboration},
  author={Shen, Leixian and Wang, Yifang and Qu, Huamin and Xie, Xing and Li, Haotian},
  journal={arXiv preprint arXiv:2510.26069},
  year={2025}
}

@article{chen2026question,
  title={VizQStudio: Iterative Visualization Literacy MCQs Design with Simulated Students},
  author={Chen, Zixin and Zeng, Yuhang and Song, Sicheng and Lin, Yanna and Xu, Xian and Qu, Huamin and Xia Meng},
  journal={arXiv preprint 	arXiv:2603.00994},
  year={2026}
}

@inproceedings{chen2025cograder,
  title={CoGrader: Transforming Instructors' Assessment of Project Reports through Collaborative LLM Integration},
  author={Chen, Zixin and Wang, Jiachen and Li, Yumeng and Li, Haobo and Shi, Chuhan and Zhang, Rong and Qu, Huamin},
  booktitle={Proceedings of the 38th Annual ACM Symposium on User Interface Software and Technology},
  pages={1--18},
  year={2025}
}

@article{chen2024stugptviz,
  title={StuGPTViz: A visual analytics approach to understand student-ChatGPT interactions},
  author={Chen, Zixin and Wang, Jiachen and Xia, Meng and Shigyo, Kento and Liu, Dingdong and Zhang, Rong and Qu, Huamin},
  journal={IEEE Transactions on Visualization and Computer Graphics},
  volume={31},
  number={1},
  pages={908--918},
  year={2024},
  publisher={IEEE}
}

@inproceedings{chen2025unmasking,
  title={Unmasking deceptive visuals: Benchmarking multimodal large language models on misleading chart question answering},
  author={Chen, Zixin and Song, Sicheng and Shum, Kashun and Lin, Yanna and Sheng, Rui and Wang, Weiqi and Qu, Huamin},
  booktitle={Proceedings of the 2025 Conference on Empirical Methods in Natural Language Processing},
  pages={13767--13800},
  year={2025}
}

\appendix









\end{document}